\documentclass[11pt]{article}
\usepackage[a4paper,margin=28mm]{geometry}
\usepackage[T1]{fontenc}
\usepackage{lmodern}
\usepackage{amsmath,amssymb,amsthm,mathtools}
\usepackage{microtype,xcolor}
\usepackage[colorlinks=true,linkcolor=blue!50!black,citecolor=blue!50!black,urlcolor=blue!50!black]{hyperref}
\hypersetup{pdftitle={Approximate Hamming Convolution via Rectangular Matrix Multiplication},
  pdfauthor={Przemysław Uznański}}

\newtheorem{theorem}{Theorem}[section]
\newtheorem{lemma}[theorem]{Lemma}
\newtheorem{proposition}[theorem]{Proposition}
\newtheorem{corollary}[theorem]{Corollary}
\theoremstyle{definition}

\theoremstyle{remark}

\newcommand{\E}{\mathbb E}
\newcommand{\Var}{\operatorname{Var}}
\newcommand{\Ham}{\operatorname{Ham}}
\newcommand{\softO}{\widetilde O}
\newcommand{\ind}[1]{\mathbf 1_{\{#1\}}}
\newcommand{\eps}{\varepsilon}
\newcommand{\ceil}[1]{\left\lceil #1\right\rceil}

\title{Approximate Hamming Convolution via Rectangular Matrix Multiplication}
\author{Przemys\l{}aw Uzna\'{n}ski\\[2pt]\small Pathway}
\date{}

\begin{document}
\maketitle

\begin{abstract}
We study $(1\pm\eps)$-approximate text-to-pattern Hamming distance:
given a length-$N$ pattern
and a length-$2N$ text, compute the Hamming distance at every alignment
to relative error $\eps$. For every fixed $a\ge0$ and
$\eps=N^{-a}$, we give a randomized algorithm with simultaneous
high-probability correctness on a logarithmic-word RAM and expected
running time
\[
 \softO\!\left(N^{\omega(4a)/2}\right),
\]
where $\softO$ suppresses subpolynomial factors and $\omega(k)$ is
the exponent for multiplying an
$n\times n^k$ matrix by an $n^k\times n$ matrix.
With rectangular bound~\cite{uznanski-2026-rectangular-polynomial-products},
this improves the published bounds~\cite{chan-golan-kociumaka-kopelowitz-porat-2020-approximating-hamming,fischer-kirkpatrick-nakos-2026-sumset-estimation}
for $1/12<a<(3+\sqrt3)/16$. In particular, the algorithm runs in
expected time $\softO(N)$ for $0\le a\le1/8$, extending the
previous near-linear guarantee through $a=1/12$.

\end{abstract}

\section{Introduction and results}
\label{sec:intro}

For equal-length strings $X,Y$ over an integer alphabet $\Sigma$,
their Hamming distance is
\[
 \Ham(X,Y)=|\{j:X[j]\ne Y[j]\}|.
\]
Given a text $T\in\Sigma^n$ and a pattern $P\in\Sigma^m$, where
$m\le n$, their \emph{text-to-pattern Hamming distance} is the array
\begin{equation}
 S[i]=\Ham\bigl(T[i:i+m],P\bigr),\qquad 0\le i\le n-m.
                                                        \label{eq:distances}
\end{equation}
We use zero-based indices and half-open substrings. For $0<\eps\le1$, a
\emph{$(1\pm\eps)$-approximation} is an array $S_\eps$ with
$(1-\eps)S[i]\le S_\eps[i]\le(1+\eps)S[i]$ for every $i$.

By the standard blocking reduction, we may assume $m=N$ and $n=2N$.
We require simultaneous correctness at all alignments with high probability.
The alphabet has size $N^{O(1)}$, and symbols occupy $O(\log N)$ bits.
Reversing the pattern expresses the equality counts as convolution
coefficients; this accounts for the term \emph{Hamming convolution}.

We use a word RAM with $\Theta(\log N)$-bit words and include all
preprocessing and output costs. Throughout, $\softO$ suppresses
subpolynomial factors in the input length, including polylogarithmic
factors. We write $\eps=N^{-a}$ for fixed $a\ge0$
when stating our running-time bounds.
Our algorithm improves the published running-time bounds for
$1/12<a<(3+\sqrt3)/16$.

\subsection{Main result}

Let $\omega(k)$ denote the arithmetic exponent for multiplying
an $n\times\ceil{n^k}$ matrix by an
$\ceil{n^k}\times n$ matrix over the rationals.
The shape parameter $k\ge0$ is fixed, and every fixed positive
exponent slack is permitted. We use the same $\omega(k)$ for word-RAM
multiplication of polynomially bounded integer matrices.
Reading the matrices and writing their product are included; the size
bound $\omega(k)\ge\max\{2,1+k\}$ covers these costs.

\begin{theorem}
\label{thm:transfer}
Fix $a\ge0$ and set $\eps=N^{-a}$.
For all sufficiently
large $N$, all distances in \eqref{eq:distances} admit simultaneous
relative-$\eps$ estimates with probability at least $1-N^{-10}$ in
expected time
\begin{equation}
 \softO\!\left(N^{\omega(4a)/2}\right).
                                                        \label{eq:transfer}
\end{equation}
The sketch overhead is polylogarithmic independently of $a$; constants
may depend on $a$ and the alphabet bound.
\end{theorem}

We reduce all alignments to a matrix product whose entries are Hamming
distances between binary rows and columns. Their supports are sets of
position--symbol pairs. Each mismatch contributes two keys to the
symmetric difference, and padding contributes a known even number.
Pair the surviving keys from the leaves of a binary tree upwards.
A node forms a pair exactly when both children have odd counts.
Sampling these nodes estimates all entries with one ordinary integer
matrix product per repetition.

\subsection{Improvement range}
\label{subsec:upper-envelope}

For the numerical consequence, we use rectangular bound
\cite[Corollary~1.2]{uznanski-2026-rectangular-polynomial-products}:
\begin{equation}
 \omega(k)\le
 \begin{cases}
 2,&0\le k\le1/2,\\
 1+k+1/(4k),&k\ge1/2.
 \end{cases}                                         \label{eq:matrix-input}
\end{equation}
These arithmetic bounds hold over fields of characteristic zero,
including the rationals.

\begin{corollary}
For $0\le a\le1/8$, all alignments can be approximated in expected time
$\softO(N)$. For $a\ge1/8$, the running time is
$\softO(N^{1/2+2a+1/(32a)})$.
\end{corollary}
\begin{proof}
Substitute $k=4a$ in \eqref{eq:matrix-input} and halve the exponent
as in Theorem~\ref{thm:transfer}.
\end{proof}

For $a\ge1/8$, the exponent in the corollary is below the
$1+2a/3$ bound of
\cite{fischer-kirkpatrick-nakos-2026-sumset-estimation} up to their
intersection at $(3+\sqrt3)/16$. If $\omega(1)=2$,
Theorem~\ref{thm:transfer} gives near-linear time through $a=1/4$.

\subsection{Previous work and the scope of the comparison}

Kopelowitz and Porat approximated all text-to-pattern Hamming distances
to relative error $\eps$ with high probability in $\softO(n/\eps)$
time~\cite[Theorem~1]{kopelowitz-porat-2015-breaking-the-variance}.
Their simpler algorithm hashes the alphabet to $O(1/\eps)$ symbols and
computes the projected distances by convolution, giving the same guarantee in
$O((n/\eps)\log n\log m)$ time~\cite[Theorem~2]{kopelowitz-porat-2018-simple-hamming-approximation}.

Chan, Golan, Kociumaka, Kopelowitz, and Porat~\cite{chan-golan-kociumaka-kopelowitz-porat-2020-approximating-hamming}
introduced structured modular sampling and the use of rectangular
matrix products for approximate Hamming matching. Their algorithm
already gives near-linear running time for sufficiently long patterns
relative to the inverse accuracy. Their stated Corollary~9.2 gives
near-linear time when $m>\eps^{-\max\{4+4/\alpha,10\}}$.
With the matrix dual-exponent bound $\alpha\ge1/2$, this guarantees
$\softO(N)$ time through $a=1/12$. Our algorithm extends the guarantee
to $a=1/8$. The comparison uses their stated dependency with this
matrix-bound substitution. Our construction also groups shifts into
rows and columns. Each sketch coordinate records the parities of two
interval counts. For small distances we use the deterministic $k$-mismatch
algorithm of Clifford, Fontaine, Porat, Sach, and Starikovskaya, which
already runs in near-linear time for $k=\sqrt N$~\cite[Theorem~1.1]{clifford-et-al-2016-k-mismatch-revisited}.
Gawrychowski and Uzna\'nski later improved the bound for larger $k$ to
$\softO(N+k\sqrt N)$ on balanced inputs~\cite[Theorem~1]{gawrychowski-uznanski-2018}.

Abrahamson~\cite{abrahamson-1987-generalized-string-matching} gives the
classical exact $\softO(N^{3/2})$ algorithm. Chan, Jin, Vassilevska
Williams, and Xu~\cite{chan-jin-williams-xu-2023} remove logarithmic
factors, obtaining an $O(N^{3/2})$ Las Vegas algorithm, and give a
uniform approximation bound with inverse-accuracy exponent
$8/(11-\omega(1))$.
Fischer, Kirkpatrick, and Nakos~\cite{fischer-kirkpatrick-nakos-2026-sumset-estimation}
give the stronger uniform bound $\softO(N\eps^{-2/3})$.
We use their approximation bound as a comparison benchmark.

To compare the approximation exponent with matrix multiplication, let
$h(a)$ be the infimum achievable running-time exponent for
relative error $\eps=N^{-a}$, with the computational model and
simultaneous success guarantee of Theorem~\ref{thm:transfer}. Our result implies
\begin{equation}
 h(a)\le\frac{\omega(4a)}2,\qquad a\ge0.
                                                        \label{eq:h-upper}
\end{equation}
The reduction in~\cite[Section~4, Corollary~13]{gawrychowski-uznanski-2018}
further implies
\begin{equation}
 h(a)\ge\frac{\omega(2a-1)}2,\qquad \frac12\le a\le1.
                                                        \label{eq:h-lower}
\end{equation}

\section{Alignments as even sets}
\label{sec:keys}

At a given alignment, we want a pattern set and a text set with one
common key per match. We group shifts into blocks so that each pattern
set serves a row of alignments and each text set serves a column.
Without loss of generality, assume that
$\sqrt N$ is an even integer.\footnote{For general $N$, use
$2\ceil{\sqrt N/2}$ as the block width and the mismatch threshold.
The number of block starts is at most the rounded width plus one.}
Write $[m]=\{0,\ldots,m-1\}$. Each shift is uniquely $i=s+u$, where
$s$ is the block start, a multiple of $\sqrt N$, and $u\in[\sqrt N]$
is the offset. All offsets in a block use the same text window of length
$M=N+\sqrt N$ starting at $s$. Append $\sqrt N$ copies of $P[0]$ to $T$,
forming $T'$, so that the last window is defined as well.

Sort the $\sigma\le3N$ distinct symbols of $P$ and $T$ and relabel them
$0,\ldots,\sigma-1$ in that order. Encode symbol $c$ at position $j$
by the key $\kappa(j,c)=j+4Nc$. All positions used below lie in $[4N]$,
so a key determines both its position and its symbol. Using positions
relative to the window, define the pattern set at offset $u$ and the
text set at block start $s$ by
\begin{align*}
 A_u &= \{\kappa(j+u,P[j]): j\in[N]\},\\
 B_s &= \{\kappa(t-s,T'[t]):s\le t<s+M\},
       \qquad s=0,\sqrt N,2\sqrt N,\ldots,\quad s\le N .
\end{align*}
At a valid shift $i=s+u\le N$, pattern position $j$ meets text position
$t=i+j$. Both keys use the position $j+u=t-s$. A match therefore
cancels from $A_u\triangle B_s$, while a mismatch contributes two keys.
The $\sqrt N$ window positions outside the aligned pattern each contribute
one text key. Thus the symmetric difference has even size and
\begin{equation}
 \frac{|A_u\triangle B_s|}{2}=S[s+u]+\frac{\sqrt N}{2}.
                                                        \label{eq:padding-error}
\end{equation}

Now take the indicator vectors of the $A_u$ as rows and those of the
$B_s$ as columns, with one coordinate for each of the $4N\sigma$ possible
keys. At a valid shift, the ordinary dot product counts the $N-S[s+u]$
matches. Summing the coordinatewise XORs gives the row--column Hamming
distance $|A_u\triangle B_s|$. Equation~\eqref{eq:padding-error} therefore
reduces all alignments to this generalized matrix product.
The $\sqrt N$ rows and $\sqrt N+1$ columns contain $\Theta(N^{3/2})$ ones
in total. Sections~\ref{sec:sketch} and~\ref{sec:batch} sketch the sets
directly from $P,T$ and estimate these distances using one ordinary
integer matrix product per repetition.

\section{Estimating the size of a set}
\label{sec:sketch}

Choose $U=2^d$ to be the smallest power of two at least $4N(\sigma+1)$.
All keys lie in $[U]$, and $d=O(\log N)$. We use the binary interval tree
on $[U]$, whose $2^j$ nodes at depth $j$ are intervals of equal size.

For a uniform odd $\mu\in[U]$, put $\pi(x)=\mu x\bmod U$. At a depth
with $p=2^j$ nodes, two distinct keys $x,y$ land in the same interval
with probability at most $2/p$ by multiply-shift hashing
\cite[Section~2.3]{thorup-2015-high-speed-hashing}. Indeed, if $g$ is
the largest power of two dividing $x-y$, then $\mu(x-y)\bmod U$ is
uniform over the $U/(2g)$ odd multiples of $g$ in $[U]$. A collision
requires a cyclic distance less than $W=U/p$. For $p\ge2$ this is
impossible if $W\le g$, and otherwise at most $W/g$ of the possible
differences qualify. The case $p=1$ is immediate.

\begin{lemma}[Set cardinality]\label{lem:accuracy}
Let $C\subseteq[U]$, where $U=2^d$ and $d\ge1$, and put $H=|C|$.
Fix $\eta>0$. Choose a uniform odd $\mu\in[U]$ and mark the leaves
$\pi(C)$, where $\pi(x)=\mu x\bmod U$. Make $L=\ceil{12d^2/\eta^2}$
independent draws using this multiplier. Draw $\ell$ chooses a uniform
depth $j_\ell\in\{0,\ldots,d-1\}$ and then a uniform node at that depth.

Let $\lambda_\ell(C)$ be the two parity bits for the numbers of marked
leaves in the left and right children of that node, respectively.
The sketch is $\operatorname{sk}(C)=(b(C),\lambda_1(C),\ldots,\lambda_L(C))$,
where $b(C)=|C|\bmod 2$. Define the cardinality estimate by
\begin{equation}
 \xi(C)=b(C)+\frac{2}{L}\sum_{\ell=1}^L d\,2^{j_\ell}
       \ind{\lambda_\ell(C)=(1,1)}.                   \label{eq:estimator}
\end{equation}
Then $\E[\xi(C)\mid\mu]=H$ for every $\mu$, and with probability at least
$2/3$ over the multiplier and the draws, $|\xi(C)-H|\le\eta H$.
\end{lemma}
\begin{proof}
Pair marked leaves from the bottom up, passing one unpaired leaf to
the parent whenever a subtree contains an odd number. A node forms one
pair exactly when both children have odd counts. The root has $b(C)$
unpaired leaves, so $(H-b(C))/2$ nodes form a pair. Conditional on $\mu$,
a node at depth $j$ is drawn with probability $1/(d\,2^j)$.
Its weight $d\,2^j$ cancels this probability. Thus the expected weighted
indicator in each draw is $(H-b(C))/2$, proving the mean identity.

Let $Z$ be the weighted indicator for one draw. Fix its depth $j$ and put
$p=2^j$. A nonzero contribution requires two marked keys in the sampled
interval. A fixed pair of keys shares an interval
with probability at most $2/p$ over $\mu$, and that interval is
selected with probability $1/p$. Hence
\[
 \E[Z^2\mid j]\le (dp)^2\frac{2\binom H2}{p^2}\le d^2H^2.
\]
The draws are independent conditional on $\mu$, and their conditional
mean is the same for every $\mu$. It follows that
$\Var \xi(C)=4\E_\mu[\Var(Z\mid\mu)]/L\le 4d^2H^2/L$.
For $H>0$, Chebyshev bounds the failure probability by
$4d^2/(L\eta^2)\le 1/3$. For $H=0$ the estimate is zero.
\end{proof}

\paragraph{Sketching a symmetric difference.}
Use the same multiplier and sampled nodes for $A$ and $B$. In every
tree interval, keys common to both sets contribute twice to the sum of
their counts. Taking parities, in the whole tree and in each sampled
child, gives the bitwise identity
\[
 \operatorname{sk}(A\triangle B)=\operatorname{sk}(A)\oplus\operatorname{sk}(B).
\]
For the application in Section~\ref{sec:keys}, $H=|A\triangle B|$
is even, so $b(A\triangle B)=0$. The XOR of two labels equals $(1,1)$
exactly when they are bitwise complements. Thus \eqref{eq:estimator} gives
\begin{equation}
 \frac{\xi(A\triangle B)}{2}=\frac{1}{L}\sum_{\ell=1}^L d\,2^{j_\ell}
 \ind{\lambda_\ell(A)=\overline{\lambda_\ell(B)}},     \label{eq:sketch-decoder}
\end{equation}
where the bar flips both bits.

\section{Batching all alignments}
\label{sec:batch}

\paragraph{Range counts for the shifted sets.}
Fix the multiplier $\mu$. Each bit of a label $\lambda_\ell(C)$ is the
parity of $|\pi(C)\cap I|$ for one of the two child intervals $I$.
The whole-set parities are known from $|A_u|=N$ and $|B_s|=N+\sqrt N$.
To answer the interval queries for every shift, put
$p_j=\pi(\kappa(j,P[j]))$ and $q_t=\pi(\kappa(t,T'[t]))$.
Shifting a position by $h$ adds $\mu h$ to its permuted key modulo $U$.
Thus, with interval translations taken modulo $U$,
\begin{equation}
 \begin{aligned}
 |\pi(A_u)\cap I| &= \bigl|\{j\in[N]:p_j\in I-\mu u\}\bigr|,\\
 |\pi(B_s)\cap I| &= \bigl|\{t\in[s,s+M):q_t\in I+\mu s\}\bigr|.
 \end{aligned}                                        \label{eq:shifted-range-counts}
\end{equation}
The pattern uses positions $j+u$, while the text window uses positions
$t-s$, which gives the two signs.

For the first count, sort the $p_j$ and find the interval endpoints by
binary search in $O(\log N)$ time. The second count is a rectangle query
on the points $(t,q_t)$. Build a balanced binary tree on $t$ and store at
each node the sorted $q_t$ values in its subtree. The window $[s,s+M)$
decomposes into $O(\log N)$ nodes, so binary searches in their lists take
$O(\log^2 N)$ time. In either query, a translated interval that wraps
around $U$ splits into two ordinary intervals. Building both structures
takes $O(N\log N)$ time and space per multiplier by sorting the pattern
keys and merging the text lists bottom up.

\begin{proposition}\label{prop:batch}
Given the shared multiplier and $L\ge1$ sampled nodes, one can generate
in $\softO(N+\sqrt N\,L)$ total time
a $\sqrt N\times 4L$ integer matrix $V$ from the row sets $A_u$ and
a $4L\times(\sqrt N+1)$ integer matrix $W$ from the column sets $B_s$.
Their product satisfies $(VW)_{u,s}/L=\xi(A_u\triangle B_s)/2$ for every $u,s$.
\end{proposition}
\begin{proof}
Process one sampled node at a time, generating four columns of $V$ and
four rows of $W$. By \eqref{eq:shifted-range-counts}, two range counts
give each two-bit label. There are $2\sqrt N+1$ labels per node, so the
preprocessing and label construction take $\softO(N+\sqrt N\,L)$ time.

Index the four coordinates of block $\ell$ by the two-bit labels.
In row $u$ put weight $d\,2^{j_\ell}$ at $\lambda_\ell(A_u)$, and in
column $s$ put weight $1$ at $\overline{\lambda_\ell(B_s)}$. Set the
other coordinates in each block to zero. Summing the block dot products
gives $L\,\xi(A_u\triangle B_s)/2$ by \eqref{eq:sketch-decoder}. The
integer weights are at most $dU$, and every product entry is at most $LdU$.
\end{proof}

\begin{proof}[Proof of Theorem~\ref{thm:transfer}]
Take $\eta=\eps/2$, so Lemma~\ref{lem:accuracy} uses
$L=\ceil{48d^2N^{2a}}$ samples. Repeat the construction of
Proposition~\ref{prop:batch} independently $O(\log N)$ times, drawing
a new multiplier each time.
Lemma~\ref{lem:accuracy}, medians and a union bound over the $N+1$
valid shifts give simultaneous relative-$\eta$ estimates of
$|A_u\triangle B_s|/2$ with failure probability at most $N^{-10}$.

On our length-$N$ pattern and length-$2N$ text, the $k$-mismatch algorithm
of Clifford et al.~\cite[Theorem~1.1]{clifford-et-al-2016-k-mismatch-revisited}
takes $\softO(N+k^2)$ time. With $k=\sqrt N$, it returns all distances
$S[i]\le\sqrt N$ exactly and identifies the rest in $\softO(N)$ time.
Keep its exact answers. For $S[i]>\sqrt N$, subtract $\sqrt N/2$ from the
median for $i=s+u$. By \eqref{eq:padding-error}, the absolute error
is at most $\eta(S[i]+\sqrt N/2)<3\eps S[i]/4$ on the simultaneous success event.

Both outer matrix dimensions are $O(\sqrt N)$. The inner dimension
is $O(N^{2a}\log^2 N)$, so splitting it into blocks of length
$\ceil{N^{2a}}$ takes only polylogarithmically many products. For fixed
$a$, the matrix indices and the product entries (bounded by $LdU$) have
$O(\log N)$ bits, so the word-RAM convention of Section~\ref{sec:intro}
applies. The products take
$\softO(N^{\omega(4a)/2})$ time. The input and output size bound
$\omega(4a)\ge\max\{2,1+4a\}$ also covers feature construction,
summing the block products, and taking medians.
\end{proof}

\section*{AI disclosure}

The results of this paper were obtained in cooperation with GPT 6.1.
Model assisted in preparing this write-up. I take full responsibility
for the contents of this paper.

\begingroup
\small
\newcommand{\etalchar}[1]{$^{#1}$}

\endgroup

\begin{thebibliography}{CGK{\etalchar{+}}20}

\bibitem[Abr87]{abrahamson-1987-generalized-string-matching}
Karl~R. Abrahamson.
\newblock Generalized string matching.
\newblock {\em SIAM Journal on Computing}, 16(6):1039--1051, 1987.
\newblock \url{https://epubs.siam.org/doi/10.1137/0216067}.

\bibitem[CGK{\etalchar{+}}20]{chan-golan-kociumaka-kopelowitz-porat-2020-approximating-hamming}
Timothy~M. Chan, Shay Golan, Tomasz Kociumaka, Tsvi Kopelowitz, and Ely Porat.
\newblock Approximating text-to-pattern {Hamming} distances, 2020.
\newblock \url{https://arxiv.org/html/2001.00211}.

\bibitem[CJVWX23]{chan-jin-williams-xu-2023}
Timothy~M. Chan, Ce~Jin, Virginia Vassilevska~Williams, and Yinzhan Xu.
\newblock Faster algorithms for text-to-pattern {Hamming} distances.
\newblock In {\em 64th IEEE Annual Symposium on Foundations of Computer Science
  (FOCS)}, pages 2188--2203, 2023.
\newblock \url{https://arxiv.org/abs/2310.13174}.

\bibitem[CFP{\etalchar{+}}16]{clifford-et-al-2016-k-mismatch-revisited}
Rapha\"el Clifford, Allyx Fontaine, Ely Porat, Benjamin Sach, and Tatiana Starikovskaya.
\newblock The $k$-mismatch problem revisited.
\newblock In {\em Proceedings of the Twenty-Seventh Annual ACM-SIAM Symposium
  on Discrete Algorithms (SODA)}, pages 2039--2052, 2016.
\newblock \url{https://doi.org/10.1137/1.9781611974331.ch142}.

\bibitem[FKN26]{fischer-kirkpatrick-nakos-2026-sumset-estimation}
Nick Fischer, Yael Kirkpatrick, and Vasileios Nakos.
\newblock Faster approximate text-to-pattern {Hamming} distances via
  near-optimal sumset estimation.
\newblock arXiv preprint arXiv:2611.05725, 2026.
\newblock \url{https://arxiv.org/abs/2611.05725}.

\bibitem[GU18]{gawrychowski-uznanski-2018}
Pawel Gawrychowski and Przemys{\l}aw Uzna\'{n}ski.
\newblock Towards unified approximate pattern matching for {Hamming} and
  {$L_1$} distance.
\newblock In {\em 45th International Colloquium on Automata, Languages, and
  Programming (ICALP 2018)}, volume 107 of {\em Leibniz International
  Proceedings in Informatics}, pages 62:1--62:13, 2018.
\newblock
  \url{https://drops.dagstuhl.de/entities/document/10.4230/LIPIcs.ICALP.2018.62}.

\bibitem[KP15]{kopelowitz-porat-2015-breaking-the-variance}
Tsvi Kopelowitz and Ely Porat.
\newblock Breaking the variance: Approximating the {Hamming} distance in
  $1/\epsilon$ time per alignment.
\newblock In {\em 56th Annual IEEE Symposium on Foundations of Computer Science
  (FOCS)}, pages 601--613, 2015.
\newblock \url{https://ieee-focs.org/FOCS-2015-Papers/8191a601.pdf}.

\bibitem[KP18]{kopelowitz-porat-2018-simple-hamming-approximation}
Tsvi Kopelowitz and Ely Porat.
\newblock A simple algorithm for approximating the text-to-pattern {Hamming} distance.
\newblock In {\em 1st Symposium on Simplicity in Algorithms (SOSA)}, volume~61
  of {\em Open Access Series in Informatics}, pages 10:1--10:5, 2018.
\newblock \url{https://doi.org/10.4230/OASIcs.SOSA.2018.10}.

\bibitem[Tho15]{thorup-2015-high-speed-hashing}
Mikkel Thorup.
\newblock High speed hashing for integers and strings, 2015.
\newblock \url{https://arxiv.org/pdf/1504.06804v9}.

\bibitem[Uzn26]{uznanski-2026-rectangular-polynomial-products}
Przemys{\l}aw Uzna\'{n}ski.
\newblock Rectangular matrix multiplication from polynomial products, 2026.
\newblock \url{https://arxiv.org/abs/2610.10494v2}.

\end{thebibliography}
\end{document}